\documentclass[prd,twocolumn,preprintnumbers,amsmath,amssymb,nofootinbib]{revtex4}
\usepackage{amsmath,amssymb}
\def\beqra{\begin{eqnarray}}
\def\eeqra{\end{eqnarray}}
\def\beq{\begin{equation}}
\def\eeq{\end{equation}}

\preprint{DESY 06 -- 211}
\allowdisplaybreaks[2]
\begin{document}
\title{Aspects of a supersymmetric Brans-Dicke theory}

\author{Riccardo Catena} 
\affiliation{Deutsches Elektronen-Syncrotron (DESY), 
\\ 22603 Hamburg, Germany \\ {\tt (catena@mail.desy.de)}}

\begin{abstract}
We consider a locally supersymmetric theory where the Planck mass is replaced by a dynamical superfield.
This model can be thought of as the Minimal Supersymmetric extension of the Brans-Dicke theory (MSBD).
The motivation that underlies this analysis  
is the research of possible connections between Dark Energy 
models based on Brans-Dicke-like theories and supersymmetric Dark Matter scenarios. 
We find that the phenomenology  associated with the MSBD model
is very different compared to the one of the original Brans-Dicke theory:
the gravitational sector does not couple to the matter sector in a universal metric way.
This feature could make the minimal supersymmetric extension of the BD idea
phenomenologically inconsistent.
\end{abstract}

\maketitle
\section{Introduction}

During the last decade of cosmological observations, the picture
of a Universe dominated by Dark Matter and Dark Energy has emerged \cite{WMAP}. 
Although a real understanding of the microscopic nature of
such cosmological components is still missing, different theories are at present
under analysis. 

Concerning Dark Matter particles, their indirectly observed 
interaction properties naturally fit those of the lightest state of
supersymmetric models with conserved $R-$parity \cite{DM}. If supersymmetry really exists, a
simple and well motivated cosmological model should include a supersymmetric
dark matter particle.

Dark Energy interaction properties are even more obscure than the Dark Matter ones.
An example of this lack of knowledge is the difficulty to explain the extremely small value of its mass scale that, 
in a phenomenologically consistent model, should be of the order of the present value of the Hubble parameter \cite{DE}.
As a consequence, direct couplings of Dark Energy to matter fields are strongly constrained
by fifth force searches \cite{Dam}.  A possible way to avoid such constraints is to work  
in the framework of Scalar-Tensor theories \cite{ST}. In these theories
the gravitational interaction is described in terms of both a metric tensor
and a scalar field. Moreover, the energy density of this extra scalar degree of freedom can be easily 
identified with Dark Energy \cite{STDE}. An interesting feature of such models is that
the gravitational sector (that also includes the Dark Energy scalar) couples to the matter 
sector in a universal metric way so that fifth force bounds are satisfied
by construction \cite{Dam}. If Dark Energy has a scalar nature, Scalar-Tensor theories provide a natural
framework to discuss its properties.

The interesting possibility to relate a Scalar-Tensor interpretation
of Dark Energy to a supersymmetric description of Dark Matter leads
to study supersymmetric extensions of Scalar-Tensor theories.
This is the topic of the present paper.
As we will see, the results of this analysis do not rely on the particular
choice of the underlying Scalar-Tensor theory; for this reason we will consider
the simplest one, {\it i.e.} the Brans-Dicke (BD) theory \cite{BD}.

In the BD theory the Planck mass 
is replaced by a dynamical scalar field. 
In this paper we consider the supersymmetric analogous of this mechanism:
we replace in the supergravity Lagrangian the Planck mass
with a chiral superfield, the ``Planck superfield''.
Such a replacement defines the ``natural'' supersymmetric extension of the 
BD theory. Let us refer to it as 
the Minimal Supersymmetric Brans-Dicke theory (MSBD) to distinguish it
from other possible approaches.
We find that, contrary to the original BD theory, in the MSBD
the gravitational sector
does not couple to the matter sector in a universal metric way. 
As a result, possible violations of the weak equivalence principle
could make the minimal supersymmetric extension of the 
BD idea phenomenologically inconsistent.
 
In spite of this conclusion, we find the subject a good laboratory
for studying realistic models of Dark Matter-Dark Energy unification. 
For instance, alternative approaches to the problem could provide a consistent scenario 
where Dark Matter and Dark Energy are identified with different components of the Planck multiplet.

The plan of this paper is as follows. 
In section \ref{not} and \ref{BDumc} we introduce notation and review 
the BD model and the concept of universal metric coupling. 
Section \ref{susyBD} is devoted to the MSBD theory;
we will specially underline the differences between its phenomenology and the one of the original BD theory.  
Section \ref{comp} is concerned with some technical details, related with the component fields formalism, 
that should make the arguments of section \ref{susyBD} more precise. 
The results are discussed in section \ref{com}.
Finally, we list in the appendix useful expressions that we used during the computations.

\section{Notation}
\label{not}

We will use in the following the same notation and conventions of \cite{Wess}. 
We list here for clarity some of them.

The superspace is described in terms of the coordinates $(y^m, \theta_{\alpha})$.
Greek indexes label two components Weyl spinors while latin indexes the components of four-vectors.
Indexes transforming under local coordinates transformations in superspace are called Einstein indexes
and are taken from the end of the alphabet, for example $(m, \,n, \, \dots)$. 
Instead, indexes transforming under local Lorentz transformations are called Lorentz indexes and
are taken from the beginning of the alphabet, for example $(a, \,b, \, \dots)$. 
The power series expansion in $\theta_{\alpha}$ of a chiral superfield $\Phi$ is given by
\beq
\Phi(y^m,\theta_{\alpha}) = A(y^m) + \sqrt{2} \,\theta^{\alpha} \chi_{\alpha}(y^{m}) + \theta^{\alpha} \theta_{\alpha} F(y^{m})
\label{ps}
\eeq  
where $A(y^{m})$ and $F(y^{m})$ are complex scalars and $\chi_{\alpha}(y^{m})$ a Weyl spinor.
We will couple matter superfields to the minimal supergravity multiplet.
This contains the vielbein $e^a_{m}$, the gravitino $\psi^a_{\alpha}$ and two auxiliary fields:
a vector $b^a$ and a scalar $M$. Finally, covariant derivatives with respect to
supergravity transformations are denoted by $\mathcal{D_{\alpha}}$, $\bar{\mathcal{D}}^{\dot{\alpha}}$ and  $\mathcal{D}_{m}$.

\section{The Brans-Dicke theory and the universal metric coupling}
\label{BDumc}

In General Relativity the coupling between gravity and matter is described by the following
Lagrangian
\beq
\mathcal{L}_{\textrm{EH}} = -\frac{1}{2} \,e M_{\textrm{Pl}}^2 \mathcal{R}  + \mathcal{L}_{\textrm{M}}[e^{a}_{m}, \Psi] \,,
\label{eh}
\eeq
where $e\equiv det(e^{a}_{m})$, $\mathcal{R}$ is the Ricci scalar and $\Psi$ symbolically
represents all matter fields involved in the theory.
In the BD approach to the gravitational interaction the Planck mass appearing in eq.~(\ref{eh}) becomes dynamical 
by means of the substitution
\beq
M_{\textrm{Pl}}^2 \Longrightarrow \varphi^{2}(y^m) \,,
\label{sp}
\eeq
where $\varphi(y^m)$ is a real scalar field.
As a consequence eq.~(\ref{eh}) is replaced by
\beqra
\mathcal{L}_{\textrm{BD}} &=& \mathcal{L}_{\varphi}[e^{a}_{m}, \varphi] + 
\mathcal{L}_{\textrm{M}}[e^{a}_{m}, \Psi] \nonumber\\
&=& - \frac{1}{2} \,e \left( \varphi^2  \, \mathcal{R}  + \omega\, \partial_{m} \varphi \partial^{m} \varphi \right) 
+ \mathcal{L}_{\textrm{M}}[e^{a}_{m}, \Psi] \,,
\label{bd}
\eeqra
where the factor $\omega$ that multiplies the kinetic term of $\varphi$ has to be tuned
to fit the post-newtonian bounds \cite{Cassini}. 
Eq.~(\ref{bd}) gives the so called ``Jordan frame'' formulation of the theory.
In this frame the BD scalar does not appear in the matter Lagrangian and particle physics is just the standard one.
The theory can be formulated in other frames related to the Jordan one by a Weyl rescaling
of the vielbein such as $e^{a}_{m} \to  e^{a}_{m}\, e^{l(\varphi)} $, where $l(\varphi)$ is some 
$\varphi$-dependent function. 
In these alternative formulations the matter Lagrangian acquires
an explicit functional dependence from $\varphi$, {\it i.e.} 
$\mathcal{L}_{\textrm{M}} = \mathcal{L_{\textrm{M}}}[e^{a}_{m} \, e^{l(\varphi)} ,\Psi]$.  
However, the inverse Weyl rescaling $e^{a}_{m} \to  e^{a}_{m}\, e^{-l(\varphi)} $
always brings back the theory to its original version in which
particle physics is just the standard one.

Eq.~(\ref{bd}) shows that in the BD theory all matter fields feel the gravitational interaction
through the same vielbein, the Jordan frame vielbein. For this reason such a matter-gravity
coupling is also called {\it universal and metric}.  
This is a non trivial property and has very important phenomenological implications.
It can be shown, for instance, that in a theory where matter couples to gravity in a universal metric
way the weak equivalence principle is satisfied by construction \cite{Dam}.

A typical example of non universal metric coupling is the following.
Let us introduce in the gravitational sector a long range scalar field $\phi$ that couples like a dilaton
to the field strength $F_{\mu\nu}$ of some (for simplicity) abelian gauge group with gauge coupling $\bar{g}$ 
\beq
S_{\phi\textrm{FF}} = 
-\frac{1}{4 \bar{g}^2} \int d^{4}x \sqrt{-g}\, g^{\alpha\mu} g^{\beta\nu} \phi \, F_{\alpha\beta} F_{\mu\nu} \, 
\eeq
where $g_{\mu\nu}$ is the metric tensor and $g\equiv det(g_{\mu\nu})$.  
Since in four dimensions the combination $\sqrt{-g}\, g^{\alpha\mu} g^{\beta\nu}$ is Weyl invariant,
the scalar field $\phi$ can not be reabsorbed by means of a rescaling of the metric. 
Therefore, in this example the gauge field
strength $F_{\mu\nu}$ feels gravity through the metric $g_{\mu\nu}$ {\it and} the scalar $\phi$. 
In other words, since no Weyl rescaling of $g_{\mu\nu}$ can ``remove'' $\phi$  
from the matter sector~$(S_{\phi\textrm{FF}})$,
it is not possible in this case to define a Jordan frame. Such a non metric and universal coupling can be easily
interpreted in terms of an effective, scalar-field dependent, gauge coupling, {\it i.e.} 
$\bar{g}^{-2}_{\textrm{eff}}(\phi) \equiv \bar{g}^{-2}\phi$. 
Moreover, it can be shown that in this picture
also the masses of the particles become $\phi$-dependent. However 
the proton and neutron masses, for instance, acquire {\it different} dependences from $\phi$. 
This is a consequence of the fact that a gauge interaction contributes differently
to the proton and neutron binding energies. As a result, the theory manifestly violates the 
weak equivalence principle~\cite{Dam}.

\section{The Minimal Supersymmetric Brans-Dicke theory}
\label{susyBD}

Eq.~(\ref{sp}) gives a prescription to construct the BD Lagrangian starting from the Einstein-Hilbert one.
In this section we apply an analogous prescription to the supergravity Lagrangian
\beq
\mathcal{L}_{\textrm{sg}} = -3\,M_{\textrm{Pl}}^{2} \int d^2\theta \, 2 \mathcal{E} R + \mathcal{L}_{\textrm{M}}[H, \Psi] + \textrm{h.c.} \,, 
\label{sg}
\eeq
where $H$ is the supergravity multiplet,
$\mathcal{E}$ is the chiral density and
$R$ represents the curvature superfield, defined as the covariant derivative of the spin connection.\\
Let us start introducing a chiral superfield $\Phi$ with components given in the power series expansion~(\ref{ps}).
We will call $\Phi$ the Planck superfield. This dynamical object allows
the natural supersymmetric extension of the substitution~(\ref{sp})
\beq
M_{\textrm{Pl}}^2 \Longrightarrow \Phi^{2}(y^m,\theta_{\alpha}) \,.
\label{sr}
\eeq

Applying the substitution~(\ref{sr}) to eq.~({\ref{sg}}) one finds 
\beqra
\mathcal{L}_{\textrm{MSBD}} &=& \mathcal{L}_{\Phi}[H, \Phi] + \mathcal{L}_\textrm{{M}}[H, \Psi] \nonumber\\ 
&=& -3 \int d^2\theta \, \Phi^2 \, 2 \mathcal{E} R - \nonumber\\
&-&\frac{1}{8} \int d^2\theta \, 2 \mathcal{E}  \left(\bar{\mathcal{D}}_{\dot{\alpha}} \bar{\mathcal{D}}^{\dot{\alpha}} 
-8\,R \right) \Phi^{\dagger} \Phi  + \nonumber\\
&+& \mathcal{L}_{M}[H,\Psi]  \,+\, \textrm{h.c.} \,, 
\label{SBD}
\eeqra
where in the third line, in  analogy with eq.~(\ref{bd}), we introduced a kinetic term for $\Phi$.
To be as general as possible we do not assume any particular form for $\mathcal{L}_{M}$.

Eq.~(\ref{SBD}) defines the Minimal Supersymmetric Brans Dicke theory (MSBD).
Its invariance under supergravity transformations follows from the properties of chiral
densities. By definitions, chiral densities transform like total derivatives
in the space $(y^{m}, \theta_{\alpha})$ and the product of a chiral density and a chiral
superfield is again a chiral density \cite{Wess}. Moreover, the superfields $\left(\bar{\mathcal{D}} \bar{\mathcal{D}} 
-8\,R \right) \Phi^{\dagger} \Phi$ and $\Phi^{2}$ are chiral if $\Phi$ is chiral.
This proves the invariance of the Lagrangian~(\ref{SBD}) under supergravity transformations.

Let us focus now on its phenomenology. As we will see explicitly in the next section, 
the component fields expansion of eq.~(\ref{SBD}) gives rise to a Lagrangian with the following structure 
\beqra
\mathcal{L}_{\textrm{MSBD}} &=& \mathcal{L}_{\Phi}[e^{a}_{m}, \psi^{a}_{\alpha}, b^{a}, M, A, \chi_{\alpha}, F]    \nonumber\\
&+& \mathcal{L}_{\textrm{\textrm{M}}}[e^{a}_{m}, \psi^{a}_{\alpha}, b^{a}, M, \Psi] \,,
\label{co}
\eeqra
where each fields was already introduced during the previous sections.
Eq.~(\ref{co}) is the supersymmetric version of eq.~(\ref{bd}).
The crucial difference between the two Lagrangians is that in the supersymmetric one
$\mathcal{L}_{\textrm{M}}$ and $\mathcal{L}_{\Phi}$ communicate 
also through the auxiliary fields $b^{a}$ and $M$. 
This has deep phenomenological consequences when the auxiliary fields are removed by means of their equations of motion.
To show this point, let us write the general solution of the equations of motion for $M$ and $b^{a}$ as follows 
\beqra
b^{a} &=& h_{1}(\dots, A,\chi_{\alpha}) \,, \nonumber\\
M &=& h_{2}(\dots, A,\chi_{\alpha}) \,,
\label{bm}
\eeqra
where $h_{1}$ and $h_{2}$ are two appropriate functions of the fields involved in the theory.
In eq.~(\ref{bm}) we underlined the crucial dependence of $h_{1}$ and $h_{2}$ from $A$ and $\chi_{\alpha}$. 
Now, replacing the solutions~(\ref{bm}) in the Lagrangian~(\ref{co}), the degrees of freedom of the
Planck multiplet explicitly appear in the matter Lagrangian. Since no Weyl rescaling of the vielbein
can remove the auxiliary fields from $\mathcal{L}_{\textrm{M}}$,  it follows that the Planck
multiplet couples intrinsically to matter. 
Therefore, there is no way to write the matter Lagrangian as 
$ \mathcal{L}_{\textrm{\textrm{M}}}[e^{a}_{m}, \psi^{a}_{\alpha}, \Psi]$ by means of a 
suitable vielbein redefinition of the form 
$e^{a}_{m} \to  e^{a}_{m}\, e^{l(A,\chi_{\alpha},F)}$, where $l$ is an appropriate function
of the components of $\Phi$.
In other words, a Jordan frame does not exist for such a theory.
The main consequence is that in the MSBD theory
the weak equivalence principle is not satisfied by construction and time variations
of masses and couplings are not under control.  
In the next section we will give the explicit expressions for eqs.~(\ref{co}) and~(\ref{bm}).

\section{Component fields}
\label{comp}

The Lagrangians given in this section are obtained using the results
summarized in the appendix. Let us start with the first term of eq.~(\ref{SBD}). 
Its component fields expansion reads 
\beqra
&&  -3 \int d^2\theta \, \Phi^2 \, 2\mathcal{E} R + \textrm{h.c.} = \nonumber\\
&& -\frac{1}{4}\,e (A^2 + A^{2*})  \mathcal{R}  \nonumber\\ 
&& + \frac{1}{2} \, e \varepsilon^{abcd} \left( \bar{\psi}_{a} \bar{\sigma}_{b} \mathcal{D}_{c}\psi_{d} A^2 -
 \psi_{a} \sigma_{b} \mathcal{D}_{c}\bar{\psi}_{d} A^{2*} \right) \nonumber\\
&&  + \frac{1}{16} \, e (A^2 - A^{2*}) \varepsilon^{abcd} \left( \bar{\psi}_{a} \bar{\sigma}_{b} \mathcal{D}_{c}\psi_{d} +
 \psi_{a} \sigma_{b} \mathcal{D}_{c}\bar{\psi}_{d} \right)  \nonumber\\
&& -\frac{1}{\sqrt{2}} \,e A \chi \sigma^{a} \bar{\sigma}^{b} \psi_{ab} 
- \frac{1}{\sqrt{2}} \,e A^{*} \bar{\chi} \bar{\sigma}^{a} \sigma^{b} \bar{\psi}_{ab}  \nonumber\\
&&- \frac{1}{6}\,e(A^2 + A^{2*}) M M^{*} +\frac{1}{6}\,e (A^2 + A^{2*})b^{a}b_{a}  \nonumber\\
&& -\frac{i}{2} \,e\,e^{m}_{a} \mathcal{D}_{m} b^{a}(A^2 - A^{2*}) \nonumber\\
&& -\frac{1}{4} \,e \psi_{a} \sigma^{a} \bar{\psi}_{b} b^{b}  (A^2 - A^{2*}) 
-\frac{1}{4} \,e \bar{\psi}_{a} \bar{\sigma}^{a} \psi_{b} b^{b}  (A^2 - A^{2*}) \nonumber\\
&& -\frac{i}{\sqrt{2}} \, e A \chi \psi_{a} b^{a} + \frac{i}{\sqrt{2}} \, e A^{*} \bar{\chi} \bar{\psi}_{a} b^{a}   \nonumber\\
&& -\frac{1}{2} \,e \chi\chi M - \frac{1}{2} \,e \bar{\chi} \bar{\chi} M^{*} + e\,AFM +e\, A^{*} F^{*} M^{*} \,,
\label{sbd}
\eeqra 
where
\beqra
&&\psi_{nm}^{\alpha} = \mathcal{D}_{n} \psi_{m}^{\alpha} - \mathcal{D}_{m} \psi_{n}^{\alpha}  \,, \nonumber\\
&&\mathcal{D}_{n} \psi_{m}^{\alpha} = \partial_{n} \psi_{m}^{\alpha} + \psi_{m}^{\beta} \omega_{n\beta}^{\,\,\,\,\,\, \alpha} \,, \nonumber 
\eeqra
and $\omega_{n\beta}^{\,\,\,\,\,\, \alpha}$ is the algebra-valued spin connection.

Now we focus on the kinetic term of the Planck superfield. Its component fields expansion is given by
\beqra
&&  -\frac{1}{8} \int d^2\theta \, 2\mathcal{E}  \left(\bar{\mathcal{D}}_{\dot{\alpha}} \bar{\mathcal{D}}^{\dot{\alpha}} 
-8\,R \right) \Phi^{\dagger} \Phi = \nonumber\\
&& + \frac{1}{6} \,e |A|^2 \mathcal{R} - e\,\partial_{m} A \partial^{m} A^{*} \nonumber\\
&& -\frac{i}{2} \,e \left( \chi \sigma^{m} \mathcal{D}_{m} \bar{\chi}+   \bar{\chi} \bar{\sigma}^{m} \mathcal{D}_{m}\chi \right) \nonumber\\
&& -\frac{1}{6} \,e |A|^2 \varepsilon^{abcd} \left( \bar{\psi}_{a} \bar{\sigma}_{b} \mathcal{D}_{c} \psi_{d} -
 \psi_{a} \sigma_{b} \mathcal{D}_{c} \bar{\psi}_{d} \right) \nonumber\\
&& + \frac{\sqrt{2}}{3} \,e \left( A^{*} \chi \sigma^{ab} \psi_{ab} + A \bar{\chi} \bar{\sigma}^{ab}\bar{\psi}_{ab}   \right) \nonumber\\
&&- \frac{\sqrt{2}}{2} \,e \left( \bar{\psi}_{a}\bar{\sigma}^{b} \sigma^{a} \bar{\chi} \partial_{b} A +  
\chi \sigma^{a} \bar{\sigma}^{b} \psi_{a} \partial_{b} A^{*}  \right) \nonumber\\
&& + \frac{1}{4} \,e  \varepsilon^{abcd} \left( A^{*} \partial_{a}A - A\partial_{a}A^{*}  \right) \psi_{b}\sigma_{c}\bar{\psi}_{d}\nonumber\\
&& -\frac{1}{9} \,e |A|^2 b_{a}b^{a} + \frac{i}{3} \,e\, b^{a} \left( A\partial_{a} A^{*} - A^{*}\partial_{a} A \right) \nonumber\\
&& -\frac{1}{6}\,e\,\chi\sigma^{a}b_{a} \bar{\chi} -i \frac{\sqrt{2}}{6}\,e b^{a}\left( A\bar{\psi}_{a}\bar{\chi} - A^{*}\psi_{a} \chi \right) \nonumber\\
&& + e\,FF^{*} +\frac{1}{9} \,e |A|^{2}|M|^{2} -\frac{1}{3}\,e MA^{*}F - \frac{1}{3}\,e M^{*}F^{*}A \nonumber\\
&& + \mathcal{L}_{\textrm{4}} \,, 
\label{kin}
\eeqra
where $\mathcal{L}_{\textrm{4}}$ includes only $4-$fermions interactions and it will be given afterwords.

Using eqs.~(\ref{sbd}) and (\ref{kin}) one can write the explicit component fields expansion of eq.~(\ref{SBD}).
For simplicity we decompose the final Lagrangian as follows
\beq
\mathcal{L}_{\textrm{MSBD}} = \mathcal{L}_{\textrm{K}} + \mathcal{L}_{\textrm{int}} + \mathcal{L}_{\textrm{4}} + \mathcal{L}_{\textrm{aux}} \,,
\label{deco}
\eeq 
where $\mathcal{L}_{\textrm{K}}$ is the Lagrangian for the kinetic terms of the fields contained in 
the Planck and supergravity multiplets,  $\mathcal{L}_{\textrm{int}}$ describes the interactions between
the Planck and supergravity multiplets not included in $\mathcal{L}_{\textrm{4}}$ and
$\mathcal{L}_{\textrm{aux}}$ is the Lagrangian for the auxiliary fields where we also absorbed $\mathcal{L}_{\textrm{M}}$.  
We list in the following their
explicit expressions.  $\mathcal{L}_{\textrm{K}}$ reads 
\beqra
 \mathcal{L}_{\textrm{K}} &=& 
 - \frac{1}{4} \,e f(A,A^{*}) \mathcal{R} - e\,\partial_{m} A \partial^{m} A^{*} \nonumber\\
 &-& \frac{i}{2} \,e \left( \chi \sigma^{m} \mathcal{D}_{m} \bar{\chi}+   \bar{\chi} \bar{\sigma}^{m} \mathcal{D}_{m}\chi \right) \nonumber\\
 &+& \,e \varepsilon^{abcd} \big[ g_{1}(A,A^{*})\bar{\psi}_{a} \bar{\sigma}_{b} \mathcal{D}_{c} \psi_{d}   \nonumber\\
 &+& g_{2}(A,A^{*}) \psi_{a} \sigma_{b} \mathcal{D}_{c} \bar{\psi}_{d} \big] \,,
\label{K}
\eeqra
where the functions $f$, $g_{1}$ and $g_{2}$ are defined as follows
\beqra
f(A,A^{*}) &=& A^2+A^{2*} -\frac{2}{3}|A|^2  \,, \nonumber\\ 
g_{1}(A,A^{*}) &=& \frac{9}{16}A^2 - \frac{1}{16} A^{2*} -\frac{1}{6}|A|^2 \,,   \nonumber\\ 
g_{2}(A,A^{*}) &=& \frac{1}{16}A^2 - \frac{9}{16} A^{2*} +\frac{1}{6}|A|^2 \,.
\eeqra

The Lagrangian $\mathcal{L}_{\textrm{int}}$ is given by
\beqra
 \mathcal{L}_{\textrm{int}} &=& 
\frac{\sqrt{2}}{3} \,e \left( A^{*} \chi \sigma^{ab} \psi_{ab} + A \bar{\chi} \bar{\sigma}^{ab}\bar{\psi}_{ab}   \right) \nonumber\\
&-& \frac{\sqrt{2}}{2} \,e \left( \bar{\psi}_{a}\bar{\sigma}^{b} \sigma^{a} \bar{\chi} \partial_{b} A +  
\chi \sigma^{a} \bar{\sigma}^{b} \psi_{a} \partial_{b} A^{*}  \right) \nonumber\\
&+& \frac{1}{4} \,e  \varepsilon^{abcd} \left( A^{*} \partial_{a}A - A\partial_{a}A^{*}  \right) \psi_{b}\sigma_{c}\bar{\psi}_{d} \nonumber\\
&-&\frac{1}{\sqrt{2}} \,e A \chi \sigma^{a} \bar{\sigma}^{b} \psi_{ab} 
- \frac{1}{\sqrt{2}} \,e A^{*} \bar{\chi} \bar{\sigma}^{a} \sigma^{b} \bar{\psi}_{ab}  \,.
\label{int}
\eeqra

The $4-$fermions interactions $\mathcal{L}_{\textrm{4}}$ read 
\beqra
\mathcal{L}_{\textrm{4}} &=&
+ \frac{1}{4}\,e\chi\sigma^{c} \bar{\sigma}^{b}\psi_{c} \bar{\psi}_{b} \bar{\chi} - i\frac{e\,\sqrt{2}}{8}
\bar{\psi}_{a} \big( \bar{\sigma}^{b}\eta^{ac}  \nonumber\\ 
&+& \bar{\sigma}^{a} \sigma^{c}\bar{\sigma}^{b} \big) \psi_{c}\bar{\psi}_{b}\bar{\chi} A \, +
\, \textrm{h.c.}  \,,
\label{4}
\eeqra
and finally,
\beqra
&&\mathcal{L}_{\textrm{aux}} = 
 \frac{1}{6} \,e f(A,A^{*}) b_{a}b^{a} + \frac{i}{3} \,e\, b^{a} \left( A\partial_{a} A^{*} - A^{*}\partial_{a} A \right) \nonumber\\
&& -\frac{1}{6}\,e\,\chi\sigma^{a}b_{a} \bar{\chi} -i \frac{\sqrt{2}}{6}\,e b^{a}\left( A\bar{\psi}_{a}\bar{\chi} - A^{*}\psi_{a} \chi \right) \nonumber\\
&& -\frac{i}{2} \,e\,e^{m}_{a} \mathcal{D}_{m} b^{a}(A^2 - A^{2*}) \nonumber\\
&& -\frac{1}{4} \,e \psi_{a} \sigma^{a} \bar{\psi}_{b} b^{b}  (A^2 - A^{2*}) 
-\frac{1}{4} \,e \bar{\psi}_{a} \bar{\sigma}^{a} \psi_{b} b^{b}  (A^2 - A^{2*}) \nonumber\\
&& -\frac{i}{\sqrt{2}} \, e A \chi \psi_{a} b^{a} + \frac{i}{\sqrt{2}} \, e A^{*} \bar{\chi} \bar{\psi}_{a} b^{a}   \nonumber\\
&& -\frac{1}{6} \,e\, f(A,A^{*})|M|^{2} + e\,FF^{*} -\frac{1}{3}\,e MA^{*}F - \frac{1}{3}\,e M^{*}F^{*}A \nonumber\\
&& -\frac{1}{2} \,e \chi\chi M - \frac{1}{2} \,e \bar{\chi} \bar{\chi} M^{*} + e\,AFM +e\, A^{*} F^{*} M^{*} \nonumber\\
&& + \mathcal{L}_{\textrm{M}}[e^{a}_{m}, \psi^{a}_{\alpha}, b^{a}, M, \Psi] \,.
\label{aux}
\eeqra 

To recover a complete analogy with eq.~(\ref{bd}) one has to perform in eq.~(\ref{deco}) a Weyl rescaling of the vielbein
in order to have a kinetic term for the graviton of the form $-1/2\,e |A|^2  \mathcal{R}$. However,
$\mathcal{L}_{\textrm{M}}$ also includes a contribution proportional to $\mathcal{R}$; as a consequence 
such a rescaling should be performed only after having specified $\mathcal{L}_{\textrm{M}}$.
Adding $\mathcal{L}_{\textrm{M}}$ to eq.~(\ref{sbd}) and taking the limit $A=A^{*}=M_{\textrm{Pl}}$ and $F=\chi=0$,
one gets the expression
\beqra
&& -\frac{1}{2} \,e M_{\textrm{Pl}}^2  \mathcal{R}  \nonumber\\ 
&& + \frac{1}{2} \,e  M_{\textrm{Pl}}^2  \varepsilon^{abcd} \left( \bar{\psi}_{a} \bar{\sigma}_{b} \mathcal{D}_{c}\psi_{d} -
 \psi_{a} \sigma_{b} \mathcal{D}_{c}\bar{\psi}_{d} \right) \nonumber\\
&&- \frac{1}{3}\,e  M_{\textrm{Pl}}^2 M\,M^{*} +\frac{1}{3} \,e  M_{\textrm{Pl}}^2 b^{a}b_{a}  \nonumber\\
&& + \mathcal{L}_{\textrm{\textrm{M}}}[e^{a}_{m}, \psi^{a}_{\alpha}, b^{a}, M, \Psi] \,.
\label{check}
\eeqra
that, in agreement with \cite{Wess}, gives the component fields expansion of the Lagrangian~(\ref{sg}).

As usual, auxiliary fields can be expressed in terms of other fields involved in the theory 
by means of their equations of motion. Using the Lagrangian~(\ref{aux}) one finds 
\beqra
&&b_{a} = -i \frac{1}{f} (A^{*}\partial_{a} A - A^{*}\partial_{a} A) + \frac{1}{2 f} \chi\sigma_{a} \bar{\chi} \nonumber\\
&& + i \frac{\sqrt{2}}{2f} (A\bar{\psi_{a}} \bar{\chi} - A^{*} \psi_{a} \chi) 
+ i \frac{3}{\sqrt{2} f} (A \chi \psi_{a} - A^{*} \bar{\chi} \bar{\psi}_{a}) \nonumber\\
&& + \frac{3}{4 f} (A^{2} - A^{2 *})  (\psi_{a}\sigma^{a} \bar{\psi}_{b} +  \bar{\psi}_{a}\bar{\sigma}^{a} \psi_{b} ) \nonumber\\
&& + i\frac{3}{2 f} \omega_{ma}^{\,\,\,\,\,\,m} - i\frac{3}{2 f} \partial_{m}[e^{m}_{a}(A^2  - A^{2*})] \nonumber\\
&& -\frac{3}{e f} \frac{\partial \mathcal{L}_{\textrm{M}}}{\partial b^{a}} \nonumber \,, \\
&& M = C_{1} \left( -3\,\bar{\chi} \bar{\chi}  + \frac{6}{e}  \frac{\partial \mathcal{L}_{\textrm{M}}}{\partial M^{*}} \right) \nonumber \,, \\
&& F^{*} = C_{2} \left( -3\,\bar{\chi} \bar{\chi}  + \frac{6}{e}  \frac{\partial \mathcal{L}_{\textrm{M}}}{\partial M^{*}} \right) \,,
\label{auxeq}
\eeqra
where
\beqra
&& C_{1}(A,A^{*}) \equiv \frac{1}{6|A|^2-A^{2}-A^{2*}} \,, \nonumber\\
&& C_{2}(A,A^{*}) \equiv C_{1}(A,A^{*}) \left( \frac{1}{3}\,A^{*} - A \right) \,. \nonumber 
\eeqra

In eqs.~(\ref{auxeq}) we omitted the dependence from $A$ and $A^{*}$ of the functions $f$, $C_{1}$ and $C_{2}$.
When $\mathcal{L}_{\textrm{\textrm{M}}}$ is specified, from eqs.~(\ref{auxeq}) one can explicitly compute 
the functions $h_{1}$ and $h_{2}$ introduced in section~\ref{susyBD}.

\section{Conclusions}
\label{com}

In this paper we have studied the minimal supersymmetric extension of the BD theory (MSBD) defined by eq.~(\ref{SBD}).
The underlying motivation was the research of possible connections between a Scalar-Tensor interpretation
of Dark Energy and a supersymmetric description of Dark Matter.  
Eq.~(\ref{SBD}) is obtained replacing the Planck mass with a chiral superfield
in the supergravity Lagrangian~(\ref{sg}). We called this extra superfield the 
Planck superfield.
Although this approach looks very natural, the resulting phenomenology
is radically different from the one of the original BD theory.
In the MSBD theory the extra degrees of freedom of the Planck superfield
intrinsically couple to matter and a Jordan frame formulation can not be achieved
through a suitable vielbein redefinition.
As a consequence, this theory does not satisfy the weak equivalence principle by construction.
This conclusion could make the minimal supersymmetric extension of the BD idea phenomenologically inconsistent.
\footnote{Here by ``inconsistent'' we mean that the weak equivalence principle is not satisfied
by construction. For any other possible inconsistency or constraint that apply to ST theories, 
see for instance \cite{Dam} and references therein.}

In spite of this result, we find that if a consistent supersymmetric Scalar-Tensor theory were constructed, 
it could provide a natural framework to achieve a Dark Matter-Dark Energy unification.
For instance, in such a scenario Dark Matter and Dark Energy could be identified
with different components of the Planck superfield. This issue is at present under analysis.

\appendix*
\section{}
We list here some useful $\theta$ expansions that we used for deriving the Lagrangians of section \ref{comp}.
Let us start with the chiral density $\mathcal{E}$. Its component fields expansion is given by \cite{Wess}
\beq
2\,\mathcal{E} = e\left[ 1 + i\theta\sigma^{a}\bar{\psi}_{a} -\theta\theta (M^{*} + \bar{\psi}_{a} \bar{\sigma}^{ab} \bar{\psi}_{b}) \right] \,.
\eeq

The curvature superfield has the following power series expansion \cite{Wess}
\beqra
 R &=& -\frac{1}{6} \Bigg\{ M + \theta \Bigg( \sigma^{a}\bar{\sigma}^{b}\psi_{ab} -i\sigma^{a} \bar{\psi}_{a} M
+ i \psi_{a}b^{a} \Bigg) \nonumber\\
 &+& \theta\theta \Bigg[ -\frac{1}{2}\mathcal{R} + i \bar{\psi}^{a} \bar{\sigma}^{b}\psi_{ab} + \frac{2}{3} MM^{*} 
+\frac{1}{3} b^{a}b_{a} \nonumber\\ 
 &-&i\,e_{a}^{\,\,\,m}\mathcal{D}_{m} b^{a} + \frac{1}{2} \bar{\psi}\bar{\psi}M 
-\frac{1}{2} \psi_{a} \sigma^{a} \bar{\psi}_{c} b^{c}  \nonumber\\ 
 &+& \frac{1}{8}\varepsilon^{abcd} \Bigg( \bar{\psi}_{a} \bar{\sigma}_{b} \psi_{cd} +
\psi_{a} \sigma_{b}\bar{\psi}_{cd} \Bigg) \Bigg]     \Bigg\}  \,.
\eeqra

Finally, the action of the chiral projector $\left(\bar{\mathcal{D}}_{\dot{\alpha}} \bar{\mathcal{D}}^{\dot{\alpha}} 
-8\,R \right)$ on the field $\Phi^{\dagger}$ is given by \cite{Wess}
\beqra
&&\left(\bar{\mathcal{D}}_{\dot{\alpha}} \bar{\mathcal{D}}^{\dot{\alpha}} -8\,R \right) \Phi^{\dagger} = \nonumber\\ 
&& -4F^{*} + \frac{4}{3}M A^{*} + \theta \Bigg[ -4i\sqrt{2} \sigma^{c} \hat{D}_{c} \bar{\chi} 
-\frac{2}{3} \sqrt{2} \sigma^{a} b_{a} \bar{\chi} \nonumber\\
&& + \frac{4}{3} A^{*}\Bigg( 2\sigma^{ab}\psi_{ab} - i\sigma^{a}\bar{\psi}_{a}M +i\psi_{a}b^{a}  \Bigg)   \Bigg] \nonumber\\
&&+\theta\theta \Bigg\{ -4\,e_{a}^{\,\,m} \mathcal{D}_{m}\hat{D}^{a}A^{*}  -\frac{8}{3} i b_{a} \hat{D}^{a} A^{*} \nonumber\\
&& -\frac{2}{3}\sqrt{2} \bar{\psi}_{ab} \bar{\sigma}^{ab}\bar{\chi} 
+ 2\sqrt{2}\bar{\psi}_{a}\hat{D}^{a} \bar{\chi} -\frac{8}{3}M^{*}F^{*} \nonumber\\
&& -\frac{2}{3} i\sqrt{2} \bar{\psi}_{a} \bar{\chi} b^{a}  +\frac{1}{3}i \sqrt{2} \bar{\psi}_{a}\bar{\sigma}^{a} \sigma^{c} \bar{\chi}b_{c} \nonumber\\
&& + \frac{4}{3} A^{*} \Bigg[ -\frac{1}{2}\mathcal{R} + i \bar{\psi}^{a} \bar{\sigma}^{b}\psi_{ab}  
+ \frac{2}{3} MM^{*} +\frac{1}{3} b^{a}b_{a} \nonumber\\
&& -i\,e_{a}^{\,\,\,m}\mathcal{D}_{m} b^{a} + \frac{1}{2} \bar{\psi}\bar{\psi}M 
-\frac{1}{2} \psi_{a} \sigma^{a} \bar{\psi}_{c} b^{c}  \nonumber\\ 
&&+ \frac{1}{8} \varepsilon^{abcd} \Bigg( \bar{\psi}_{a} \bar{\sigma}_{b} \psi_{cd} +
\psi_{a} \sigma_{b}\bar{\psi}_{cd} \Bigg)  \Bigg] \Bigg\} \,,
\eeqra
where
\beqra
\hat{D}_{a}A^{*} &=& e_{a}^{\,\,m} \partial_{m}A^{*} - \frac{1}{2} \sqrt{2} \bar{\psi}_{a\dot{\alpha}} \bar{\chi}^{\dot{\alpha}} \nonumber\\
\hat{D}_{a}\bar{\chi}^{\dot{\alpha}} &=& e_{a}^{\,\,m} \mathcal{D}_{m}\bar{\chi}^{\dot{\alpha}} -
\frac{i}{2} \sqrt{2} \bar{\sigma}^{b\dot{\alpha}\delta}\psi_{a\delta}\hat{D}_{b}A^{*} \nonumber\\
&-& \frac{1}{2} \sqrt{2} \bar{\psi}_{a}^{\dot{\alpha}} F^{*} \,.
\eeqra

\acknowledgments 
I sincerely thank Massimo Pietroni for many useful 
suggestions and discussions on Scalar-Tensor theories and their possible supersymmetric 
extensions. I would also like to thank Wilfried Buchmueller for
interesting discussions on the topic and Massimo Pietroni and Gonzalo Palma 
for having read and commented on a draft of the paper.
I finally acknowledges a Research Grant funded by the VIPAC Institute.


\end{document}